\documentclass[10pt,final,journal,a4paper,oneside,twocolumn]{IEEEtran}

\usepackage{bm}
\usepackage[usenames,dvipsnames]{xcolor}
\usepackage{srcltx}
\usepackage{psfrag}
\usepackage{epsfig}
\usepackage{graphicx}
\usepackage{epstopdf}
\usepackage{color}
\usepackage[tbtags,sumlimits,nointlimits,reqno]{amsmath}
\usepackage{amssymb}
\usepackage{showkeys}   
\usepackage{color}
\usepackage{float}
\usepackage{subfigure}
\usepackage{gensymb}
\begin{document}

\title{Extending the Brewster Effect to Arbitrary Angle
 and Polarization using Bianisotropic Metasurfaces}

\author{\IEEEauthorblockN{Guillaume~Lavigne and Christophe~Caloz \\}
\IEEEauthorblockA{Department of Electrical Engineering, Polytechnique Montr\'{e}al\\
 Montr\'{e}al, Qu\'{e}bec, Canada}
}
\maketitle

\begin{abstract}
A bianisotropic metasurface design is proposed for extending the Brewster effect to arbitrary angles and polarizations. The metasurface is synthesized using the surface susceptibility tensor and Generalized Sheet Transition Conditions (GSTCs) synthesis method, and is demonstrated by GSTC-FDFD simulation. It is found that an extremely broad angular range ($0-30^\circ$) with near zero reflection may be obtained near the normal angle, which is of paramount interest in paraxial optics, while narrower angular range at more oblique angles may find applications in spatial filtering.

\end{abstract}

\section{Introduction}\label{sec:intro}

The electromagnetic problem of a conventional interface between two media is described by Snell law and Fresnel coefficients~\cite{rothwell2008electromagnetics}. In the case of the interface between two \emph{purely dielectric} ($\mu_\text{r}=1$) media and for the \emph{TM (or p)} polarization, it admits an incidence angle with zero reflection, called the Brewster angle, which reads $\theta_\text{B}=\arctan(n_2/n_1)$, where $n_1$ and $n_2$ are the refractive indices of the incident and transmitted media, respectively.

Unfortunately, such a Brewster angle does not exist for the  \emph{TE (or s)} polarization\footnote{This is unless the media were magnetic.}, which prevents unpolarized light to be completed transmitted in conventional optical systems. Recently, it has been shown the both TM and TE Brewster angles could be achieved using a metasurface placed at the interface between the two dielectric media~\cite{paniagua2016generalized}. However, the metasurface in~\cite{paniagua2016generalized} does allow to possibility of having arbitrary and/or coincident TM and TE Brewster angles.

Here, we show that using a properly designed \emph{bianisotropic} metasurface~\cite{lavigne2017refraction}, these restrictions can be completely lifted. Specifically, we propose a bianisotropic metasurface design, based on the surface susceptibility tensor and Generalized Sheet Transition Conditions (GSTCs) synthesis method, that allows one to realize \emph{arbitrary and independent TM and TE Brewster angles} between two dielectric media, which is of considerable practical interest.

\section{General Sheet Transition Conditions (GSTCs)}\label{sec:gstc}

The GSTCs of a bianisotropic metasurface, that we shall assume here to purely transverse for simplicity, read~\cite{lavigne2017refraction}

\begin{subequations}\label{eq:GSTC}
\begin{equation}
\hat{z} \times \Delta\mathbf{H} = j \omega \epsilon \overline{\overline{ \chi}}_\text{ee} \mathbf{E}_\text{av} +  j \omega \overline{\overline{ \chi}}_\text{em} \sqrt{\mu \epsilon}  \mathbf{H}_\text{av} ,
\end{equation}
\begin{equation}
\Delta \mathbf{E} \times \hat{z}   = j \omega \mu \overline{\overline{ \chi}}_\text{mm} \mathbf{H}_\text{av} +  j \omega \mu \overline{\overline{ \chi}}_\text{me} \sqrt{\frac{\epsilon}{\mu}}  \mathbf{E}_\text{av} ,
\end{equation}
\end{subequations}
where the $\Delta$ symbol and the 'av' subscript represent the differences and averages of the tangential fields on both sides of the metasurface, and $\overline{\overline{ \chi}}_\text{ee}, \overline{\overline{ \chi}}_\text{em}, \overline{\overline{ \chi}}_\text{me}, \overline{\overline{ \chi}}_\text{mm}$ are the bianisotropic surface susceptibility tensors describing the metasurface.

\section{Fields Specifications}\label{sec:fields}

The synthesis procedure consists in determining the surface susceptibility tensors corresponding to specified fields with the susceptibility-GSTC equations~\eqref{eq:GSTC}. So, we now have to determine the field specifications for arbitrary angle and polarization Brewster effect. We assume here that no gyrotropy is desired and consider therefore a non-gyrotropic metasurface.
\begin{figure}[h]
\centering
\vspace{-1.2cm}
\includegraphics[width=0.9\columnwidth]{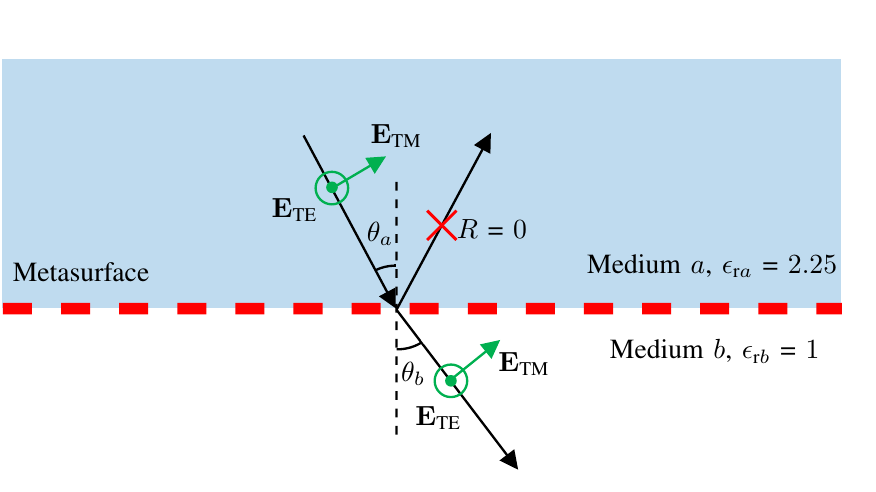}
\vspace{-0.3cm}
\caption{Design of a metasurface producing the Brewster effect for arbitrary angle ($\theta_a$) and polarization (TM and TE) between two dielectric media. In the numerical illustrations, we shall consider an air -- glass ($\epsilon_{\text{r}a} = 2.25$) interface, for which $\theta_\text{B}=33.69^\circ$.}\label{fig:Brewster_problem}
\end{figure}

The tangential fields for a TM plane wave incident in medium~$a$ at an angle $\theta_a$ and refracted without reflection (Brewster angle specification) in medium~$b$ at an angle $\theta_b$ are
\begin{subequations}\label{eq:fields_TM}
\begin{equation}
\mathbf{E}_{\parallel a,\text{TM}}=\cos\theta_a\hat{x}e^{-jk_{ax}x},\quad
\mathbf{H}_{\parallel a,\text{TM}}=\frac{e^{-jk_{ax}x}}{\eta_a}\hat{y},
\end{equation}
\begin{equation}
\mathbf{E}_{\parallel b,\text{TM}}=T\cos\theta_b\hat{x}e^{-jk_{bx}x},\quad
\mathbf{H}_{\parallel b,\text{TM}}=T\frac{e^{-jk_{bx}x}}{\eta_b}\hat{y}.
\end{equation}
\end{subequations}
Similarly, the incident and transmitted TE plane waves read
\begin{subequations}\label{eq:fields_TE}
\begin{equation}
\mathbf{E}_{\parallel a,\text{TE}}=-e^{-jk_{ax}x}\hat{y},\quad
\mathbf{H}_{\parallel a,\text{TE}}=\cos\theta_a\hat{x}\frac{e^{-jk_{ax}x}}{\eta_a},
\end{equation}
\begin{equation}
\mathbf{E}_{\parallel b,\text{TE}}=-Te^{-jk_{bx}x}\hat{y},\quad
\mathbf{H}_{\parallel b,\text{TE}}=T\cos\theta_b\hat{x}\frac{e^{-jk_{bx}x}}{\eta_b},
\end{equation}
\end{subequations}
%
where $\theta_b = \arcsin (\frac{n_a}{n_b}\sin \theta_a) $ according to Snell law, $k_{(a,b)}$ is the wavenumber and $\eta_{(a,b)}$ is the wave impedance in medium $(a,b)$. The transmission coefficient is obtained as $T =\sqrt{(\eta_b\cos\theta_a)/(\eta_a\cos\theta_b)}$ by enforcing power conservation across the metasurface with zero reflection.

\section{Bianisotropic Metasurface Synthesis}\label{sec:biani}

We consider \emph{bianisotropic} metasurface, recently shown to be required for generalized refraction~\cite{lavigne2017refraction} and described by the susceptibility components $\chi^{xx}_\text{ee}$, $\chi^{yy}_\text{mm}$, $\chi^{xy}_\text{em}$ and $\chi^{yx}_\text{me}$ for TM polarization and by $\chi^{yy}_\text{ee}$, $\chi^{xx}_\text{mm}$, $\chi^{yx}_\text{em}$ and $\chi^{xy}_\text{me}$ for TE polarization. This metasurface has 4 independent unknown susceptibility components for each polarization while Eqs.~\ref{eq:GSTC} represent only 2 equations. In order to obtain a full-rank system, we will need to specify \emph{two} transformations from~\ref{eq:GSTC} for each polarization. In order to avoid the practical complexity associated with nonreciprocity, we enforce here the reciprocity condition $\overline{\overline{\chi}}_{\text{em}}=-\overline{\overline{\chi}}^\dag_{\text{me}}$, which provides the second transformation specification. The first transformation (subscript '1' below) corresponds to the fields propagating from mediu $a$ to medium $b$, while the second transformation (subscript '2' below) corresponds to the fields propagating in the reverse direction, from medium $b$ to medium $a$ . The resulting systems of equation  can be written as
\begin{equation}\label{eq:TM_transf_syst}\hspace{-5mm}\vspace{-2mm}
\begin{bmatrix}
    \Delta H_{y1}    & \hspace{-3mm}     \Delta H_{y2} \\
    \Delta E_{x1}     & \hspace{-3mm}     \Delta E_{x2} \\
\end{bmatrix}
=
\begin{bmatrix}
    -j\omega\epsilon_0\chi^{xx}_\text{ee} &\hspace{-3mm} -jk_0\chi^{xy}_\text{em}  \\
    -jk_0\chi^{yx}_\text{me} &\hspace{-3mm} -j\omega\mu_0\chi^{yy}_\text{mm} \\
\end{bmatrix}
\begin{bmatrix}
          E_{x1,\text{av}} &\hspace{-3mm} E_{x2,\text{av}} \\
      H_{y1,\text{av}} &\hspace{-3mm} H_{y2,\text{av}}
\end{bmatrix},
\end{equation}
for TM polarization, and
\begin{equation}\label{eq:TE_transf_syst}
  \hspace{-5mm}
\begin{bmatrix}
    \Delta H_{x1}    & \hspace{-3mm}     \Delta H_{x2} \\
    \Delta E_{y1}     & \hspace{-3mm}     \Delta E_{y2} \\
\end{bmatrix}
=
\begin{bmatrix}
    -j\omega\epsilon_0\chi^{yy}_\text{ee}      &\hspace{-3mm} -jk_0\chi^{yx}_\text{em}  \\
    -jk_0\chi^{xy}_\text{me}       &\hspace{-3mm} -j\omega\mu_0\chi^{xx}_\text{mm} \\
\end{bmatrix}
\begin{bmatrix}
          E_{y1,\text{av}}      &\hspace{-3mm}   E_{y2,\text{av}} \\
      H_{x1,\text{av}} &\hspace{-3mm} H_{x2,\text{av}}
\end{bmatrix},
\end{equation}
for TE polarization, and can be solved analytically for the susceptibility as a function of
the specified Brewster angle $\theta_a$.

\begin{figure}[h]
\centering
\includegraphics[width=1\columnwidth]{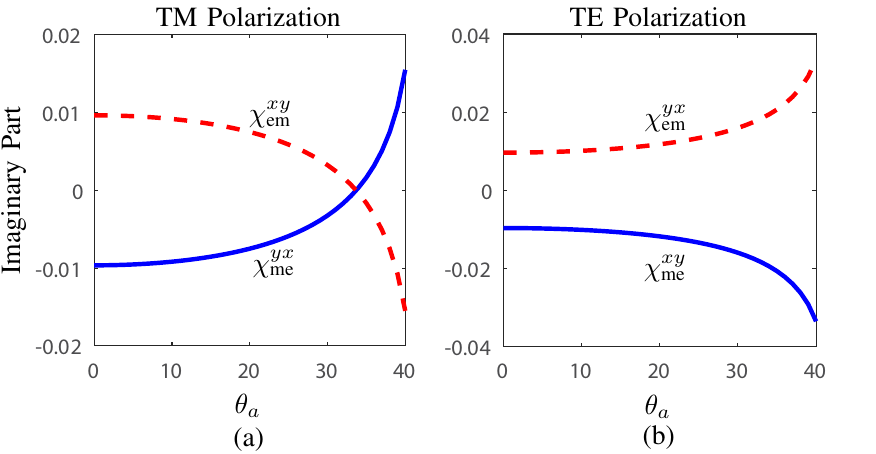}
\caption{Metasurface susceptibilities synthesized for Brewster effect at the angle $\theta_a$ for the parameters Fig~\ref{fig:Brewster_problem}. (a)~TM polarization. (b)~TE polarization.}\label{fig:fct_Chi}
\end{figure}

The resulting susceptibilities are plotted in Fig.~\ref{fig:fct_Chi} for the air -- glass interface. Interesting, the synthesis reveals that the monoanisotropic components vanish ($\chi^{yy}_\text{ee}= \chi^{xx}_\text{ee} = \chi^{yy}_\text{mm} = \chi^{xx}_\text{mm} =0$), which leads to a purely bianisotropic, passive and reciprocal metasurface. Moreover, the metasurface is \emph{uniform}, because it follows conventional Snell law in transmission. As may be expected, the susceptibilities reduce to zero for for TM polarization at the conventional Brewster angle ($\theta_B=33.69^\circ$).

A possible issue with the physical implementation of this metasurface is that a purely bianisotropic scattering particle does not theoretically exist, as shown in~\cite{albooyeh2016purely}. We propose two practical solutions: 1)~using scattering particles with $\chi_\text{ee} \approx \chi_\text{mm} \approx 0$ but not exactly zero, approximating this design, and 2)~using a combination of scattering particles providing the appropriate $\chi_\text{em}$ and $\chi_\text{me}$ while canceling each other's $\chi_\text{ee}$ and $\chi_\text{mm}$.

\section{Numerical Characterisation}\label{sec:characterisation}

\begin{figure}[h]
\centering
\includegraphics[width=0.9\columnwidth]{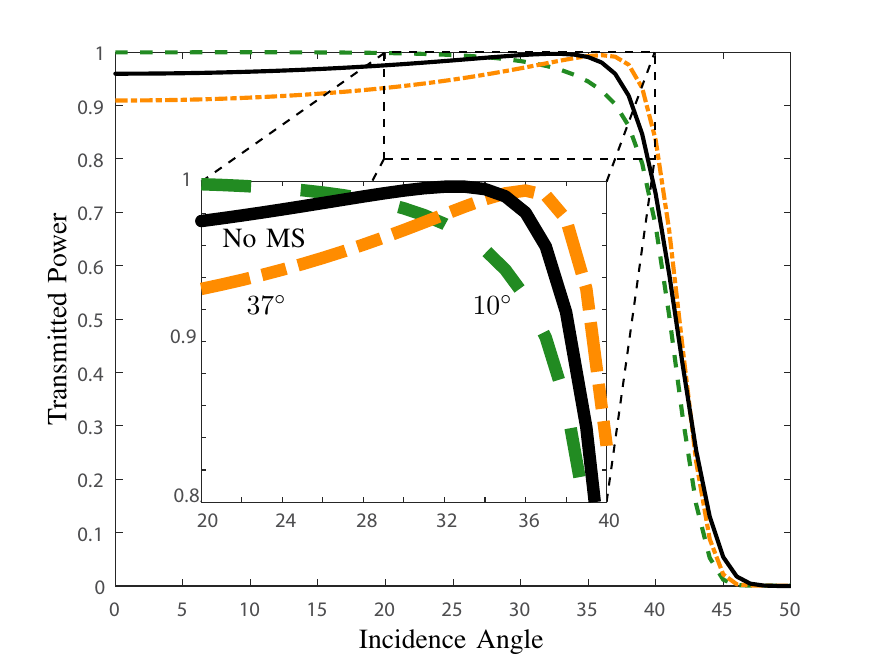}
\caption{Transmitted power as a function of the incident angle for three different cases. All the curves are symmetric (not shown here) with respect to normal incidence.}\label{fig:Brewster_plot}
\end{figure}

Figure~\ref{fig:Brewster_plot} plots the angular response, obtained by GSTC-based FDFD simulations~\cite{vahabzadeh2016simulation}, of the conventional dielectric interface (which corresponds to Fresnel coefficients) and the interface hosting two different metasurfaces with Brewster angles $\theta_a = 10^\circ$ and $\theta_a = 37^\circ$ for TM polarization. According to specification, full transmission is achieved at the specified angles for both metasurface interfaces and the transmission is also altered for all other angles of incidence, deviating from conventional Fresnel coefficients.
The metasurface with a specified Brewster of $10^\circ$ exhibits almost full transmission over a large angular spectrum around normal incidence, which is of great interest for paraxial optics. In contrast, the metasurface with Brewster $37^\circ$ has a narrow range of high transmission, which may find applications in optical spatial filters.
\vspace{-2mm}
\section{Conclusion}\label{sec:conclusion}

A bianisotropic metasurface design has been proposed for extending the Brewster effect to arbitrary angles and polarizations. The next step of the work will consist in designing the practical scattering particles and experimentally demonstrating the concept. This Brewster effect extension may find various applications in microwave and optical systems.

\bibliographystyle{IEEEtran}
\bibliography{LIB}

\end{document}